\documentclass[aps,prl,twocolumn,groupedaddress,nofootinbib,superscriptaddress,notitlepage,floatfix]{revtex4-1}
\usepackage{graphicx,graphics,subfigure}
\usepackage{dcolumn}   
\usepackage{array}
\usepackage{nicefrac}
\usepackage{amsthm,amsmath,amssymb,amsfonts,mathrsfs} 
\usepackage{bm}
\usepackage{braket} 
\usepackage{enumitem}
\usepackage{url}
\usepackage[pdfstartview=FitH,pdfencoding=auto]{hyperref}
\usepackage{bookmark}
\usepackage{pifont}
\hypersetup{
    colorlinks=true,        		
    linkcolor=blue,              	
    citecolor=blue,        			
    filecolor=blue,          		
    urlcolor=blue,              	
    runcolor=cyan
			}
\usepackage[pdftex]{color}
\usepackage{multirow}
\usepackage{CJK}
\newcommand{\tr}{\mathrm{tr}}

\begin{document}
\begin{CJK*}{GB}{}
\title{Demonstration of irreversibility and dissipation relation of thermodynamics with a superconducting qubit}
\author{Xue-Yi Guo}
	\affiliation{Institute of Physics, Chinese Academy of Sciences, Beijing 100190, China}
\affiliation{University of Chinese Academy of Sciences, Beijing 100049, China}
\thanks{X. Y. G. and Y. P. contributed equally to this work.}
\author{Yi Peng}
	\affiliation{Institute of Physics, Chinese Academy of Sciences, Beijing 100190, China}
\affiliation{University of Chinese Academy of Sciences, Beijing 100049, China}
\thanks{X. Y. G. and Y. P. contributed equally to this work.}

\author{Changnan Peng}
\affiliation{California Institute of Technology, Pasadena, CA 91125, USA}

\author{Hui Deng}
	\affiliation{University of Science and Technology of China, Hefei, Anhui 230026, China}
\affiliation{Synergetic Innovation Centre in Quantum Information and Quantum Physics,
University of Science and Technology of China, Hefei, Anhui 230026, China}

\author{Yi-Rong Jin}
\email{jyr-king@iphy.ac.cn}
	\affiliation{Institute of Physics, Chinese Academy of Sciences, Beijing 100190, China}

\author{Chengchun Tang}
	\affiliation{Institute of Physics, Chinese Academy of Sciences, Beijing 100190, China}

\author{Xiaobo Zhu}
\affiliation{University of Science and Technology of China, Hefei, Anhui 230026, China}
\affiliation{Synergetic Innovation Centre in Quantum Information and Quantum Physics,
University of Science and Technology of China, Hefei, Anhui 230026, China}

\author{Dongning Zheng}
\email{dzheng@iphy.ac.cn}
	\affiliation{Institute of Physics, Chinese Academy of Sciences, Beijing 100190, China}
\affiliation{University of Chinese Academy of Sciences, Beijing 100049, China}

\author{Heng Fan}
\email{hfan@iphy.ac.cn}
	\affiliation{Institute of Physics, Chinese Academy of Sciences, Beijing 100190, China}
\affiliation{University of Chinese Academy of Sciences, Beijing 100049, China}

\date{\today}
\eid{identifier}
\pacs{xxx}
\begin{abstract}
	We investigate experimentally the relation between thermodynamical irreversibility and dissipation on a superconducting Xmon qubit.
This relation also implies the second law and the Landauer principle on dissipation in irreversible computations.
In our experiment, the qubit is initialized to states according to the Gibbs distribution.
Work injection and extraction processes are conducted through two kinds of
	unitary driving protocols, for both a forward process and its corresponding mirror reverses. The relative entropy and relative
	R\'enyi entropy are employed to measure the asymmetry between paired forward and backward work injection or extraction processes.
	We show experimentally that the relative entropy and relative R\'enyi entropy measured irreversibility are related to the average of
	work dissipation and average of exponentiated work dissipation respectively.
Our work provides solid experimental support for the theory of quantum thermodynamics.
\end{abstract}
\maketitle
\end{CJK*}
\emph{Introduction.}---Time-reversal symmetry in the microscopic description of nature and the arrow of time in the macroscopic world described
	by thermodynamics are central topics in physics.
The theory of thermodynamics indicates that the irreversibility of a thermal process is connected to
	the work dissipation $\braket{W_\mathrm{diss}}=\braket{W}-\Delta{F}$ in transition, where $\Delta{F}$ is free energy
	difference between the final and initial equilibrium states, while $W$ is the work done on the system in the transformation. The second
	law of thermodynamics states that $\braket{W_\mathrm{diss}}{\ge}0$ for any transition process from an initial equilibrium state to
	a final equilibrium state. Thus $\Delta{F}$ is the minimum possible work that the system should absorb in the process. When the
	transition goes sufficiently slow to achieve a quasi-static transformation, the whole process would be reversible.
In this case, the external agents need only inject a necessary work $\Delta{F}$ to the system, correspondingly, dissipated work will vanish.
On the other hand, the transitions which usually take finite time are accompanied with non-zero dissipated work.
	Qualitatively, with more work dissipated in the transition, the degree of irreversibility of the whole process
	will be higher~\cite{Landau1980,Kittel1980}.

Generally, a quantitative description of the relation between irreversibility and dissipation is elusive,
which may involve non-equilibrium dynamics at the microscopic level. In this Letter, we report an experiment
on testing a relation, which will be presented later in Eq.(\ref{time_assymetry_relativeEntr}),
for irreversibility and dissipation of thermodynamics with a superconducting qubit.
We remark that this relation also implies the second law and the Landauer principle \cite{kawai2007dissipation,Maruyama,Quan1,Quan2}.
It should be noted that many experimental explorations of thermodynamics, especially involving nonequilibrium processes,
have already been conducted, mainly in classical systems such as macromoleculars~\cite{liphardt2002,collin2005verification,harris2007}, macroscopic mechanical
	systems~\cite{douarche2005}, Brownian particles~\cite{andrieux2007entropy,andrieux2008,roldan2013universal}, silica nanoparticles~\cite{gieseler2014dynamic},
	colloidal particles~\cite{blickle2006,tusch2014energy,berut2015information},
	RC circuit~\cite{andrieux2007entropy,andrieux2008,granger2015fluctuation} etc.
A few experiments are implemented in quantum systems such as nuclear magnetic resonance
	(NMR) system~\cite{Batalhao2014,batalhao2015irreversibility}. There is also a proposal for
 trapped ions~\cite{Huber2008}.

 Experimentally, with a trapped-ion system, the Jarzynski equality \cite{Jarzynski97} which
relates the free-energy difference to the exponential average work done on the system is
tested where the technique challenging method is to obtain the work distribution by two
projective measurements over energy eigenstates \cite{An2014}. Remarkably, the time asymmetry relation~(\ref{time_assymetry_relativeEntr}) under
consideration was tested in NMR system~\cite{batalhao2015irreversibility}.
However, for relation (\ref{time_assymetry_relativeEntr}),
the involved dissipation work implementation defining as energy difference for single-shot projective measurement
bases as that in \cite{An2014} is still absent, which would put this time asymmetry relation on more solid experimental foundation.
Here, we report an experimental testing of the relation (\ref{time_assymetry_relativeEntr}) and also its
relative R\'enyi entropy generalization with a qubit of superconducting quantum circuit by quantum non-demolition (QND) measurement.
This QND measurement is similar as single-shot measurement in definition but is different from
that of NMR ensemble quantum system. We know that the superconducting quantum circuit
is promising for quantum information processing for its well addressibility and controllability\cite{zheng2017,Yale2016,Barends2016,HHWang1,HHWang2,You1}.

\emph{Theoretical background.}---The relation between irreversibility and dissipation is in general intriguing,
	because processes of work extraction and heat exchanging are often intertwined together making it hard to separate them apart.
	This predicament can be resolved by arranging these two processes taking place separately in different time periods. Specifically,
	we consider a controllable system and a thermal reservoir at constant inverse temperature
	$\beta{\equiv}1/T$ with $T$ the temperature and Boltzmann constant set to 1. System is initialized at an equilibrium state in
	contacting with the heat reservoir. Then it is separated from the reservoir, and driven out of equilibrium by a unitary evolution.
	Finally, the system contacts with the reservoir again and thermalize back to an equilibrium state.
	This constitutes the forward process. The backward counterpart starts from the former ``final equilibrium state'', undergoes
	reversed unitary driven evolution and ends at the former ``initial equilibrium state'' through heat exchange with reservoir.
	Since there is no other restriction on the unitary driven process, the system can be far away from equilibrium in the
	transition, often out of the linear response regime~\cite{Kubo1957}. Such process will result in non-vanishing dissipated work
	$\braket{W_\mathrm{diss}}$ and thus irreversibility in thermodynamics. Quantitatively, a relation between time
	irreversibility and dissipation of work can be written as~\cite{kawai2007dissipation,parrondo2009entropy},
	\begin{equation}
		S\left(\pmb{\rho}^\mathrm{F}\left(t\right)\|\pmb{\rho}^\mathrm{B}\left(\tau-t\right)\right)
		= \beta\braket{W_\mathrm{diss}},
		\label{time_assymetry_relativeEntr}
	\end{equation}
	where $\tau$ is the total driven time, $\pmb{\rho}^\mathrm{F}\left(t\right)$ is state of system at time $t$ in the forward
	process, $\pmb{\rho}^\mathrm{B}\left(\tau-t\right)$ is state at time $\tau-t$ in the backward process, and
	$S\left(\pmb{\rho}^\mathrm{F}\left(t\right)\|\pmb{\rho}^\mathrm{B}\left(t\right)\right)
	                                       \equiv \tr\left\{\pmb{\rho}^\mathrm{F}\left(t\right)
												  \left\lbrack{\ln\pmb{\rho}^\mathrm{F}\left(t\right)
												  -\ln\pmb{\rho}^\mathrm{B}\left(\tau-t\right)}
												     \right\rbrack\right\}$ denotes relative entropy between the pair of states. Given
	reversible processes, states $\pmb{\rho}^\mathrm{F}\left(t\right)$ and
	$\pmb{\rho}^\mathrm{B}\left(\tau-t\right)$ coincide which is the sufficient and necessary condition that relative entropy
	between them vanishes for arbitrary time points $t$ and $\tau-t$. Relative entropy is non-negative and a
	measure of how different $\pmb{\rho}^\mathrm{F}\left(t\right)$ is from $\pmb{\rho}^\mathrm{B}\left(\tau-t\right)$. Thus it can
	quantify the asymmetry between forward and backward processes.

	Alternatively, $\alpha $-order relative R\'enyi entropy,
	$S_\alpha(\pmb{\rho}^\mathrm{F}(t)\|\pmb{\rho}^\mathrm{B}(\tau-t))
	{\equiv}\frac{1}{\alpha-1}
	        \ln\{\tr(\lbrack{\pmb{\rho}^\mathrm{F}\left(t\right)}\rbrack^\alpha
									\lbrack{\pmb{\rho}^\mathrm{B}(\tau-t)}\rbrack^{1-\alpha})\}$ with $\alpha{>}0$,
can also be used in addition to relative entropy to measure asymmetry between forward and backward processes. Similar form of thermodynamical irreversibility and
	dissipation relation can be obtained~\cite{wei2017relations},
	\begin{equation}
		S_\alpha\left(\pmb{\rho}^\mathrm{F}\left(t\right)\|\pmb{\rho}^\mathrm{B}\left(\tau-t\right)\right)
		=\frac{1}{\alpha-1}\ln\Braket{e^{\beta(\alpha-1){W_\mathrm{diss}}}}.
		\label{time_assymetry_relatiRenyiEntropy}
	\end{equation}
	As $\alpha$ approaches 1, relative R\'enyi entropy converges to relative entropy and thus is a generalization of
	the latter providing insights from different viewpoints. They are both useful tools in quantum
	information and thermodynamics~\cite{nielsen2010quantum,Vidal2002,Osborne2002,reeb2014improved,roldan2014universal,Brandao2015a}.

\begin{figure}[ht!]
	\includegraphics[height=0.50\textwidth]{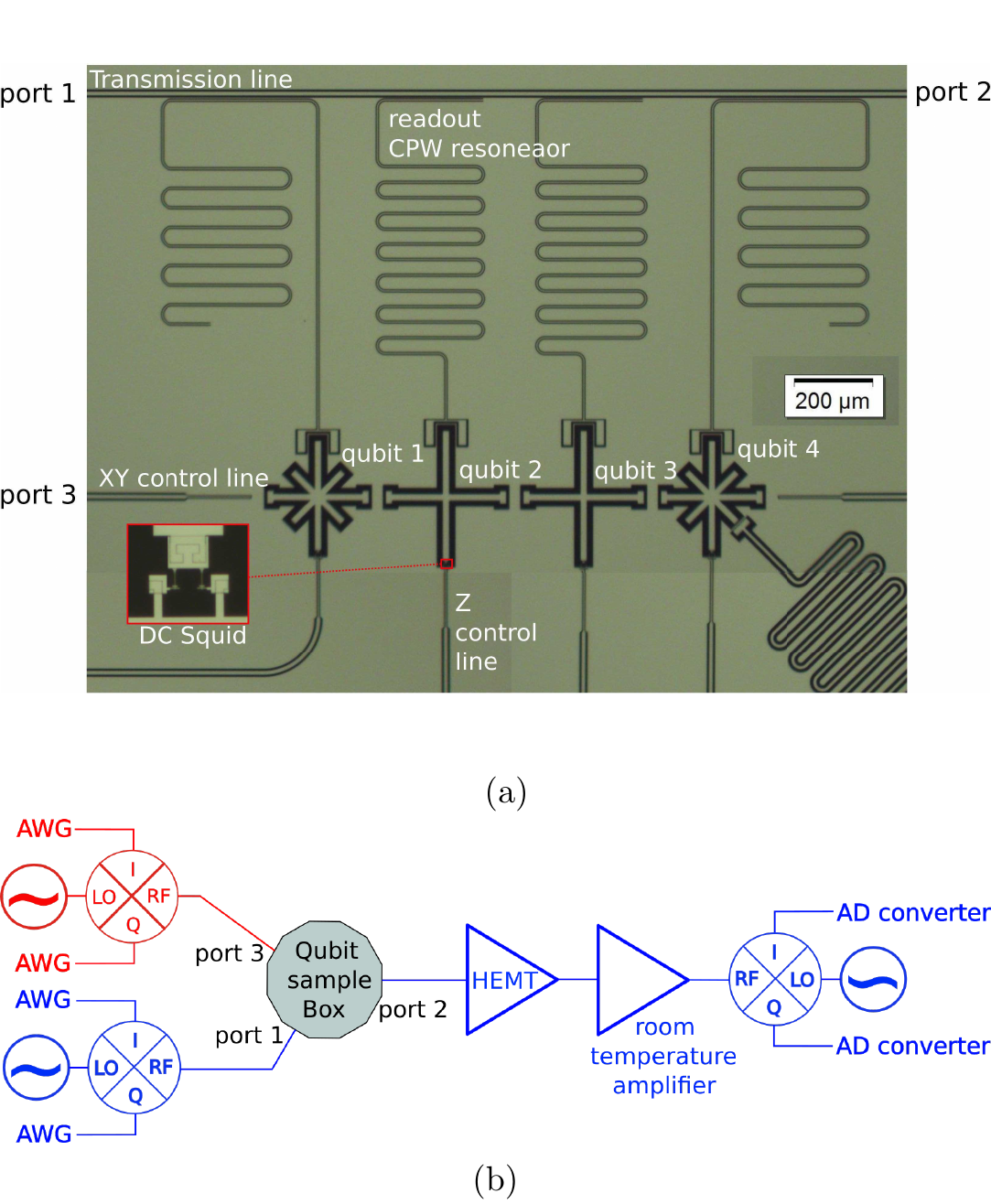}
	\caption{(Color online) (a) shows an optical micrograph of the superconducting quantum circuit used for our experiment. Four Xmon qubits (marked from "qubit 1" to "qubit 4") are arranged in a line with nearest-neighbor coupling scheme. Each qubit have its own Z control lines, while qubit 2 and 3 share their X/Y control lines with qubit 1 and qubit 4, respectively. The device was fabricated following the procedure outlined in \cite{zheng2017} with similar circuit layout. It was then mounted in a  light-tight aluminium alloy box and cooled down to a base temperature of about 10 mK. (b) shows a schematic of the qubit control and readout circuits and electronics. An arbitrary waveform generator (Tektronix AWG 5014C) was used to generate well-shaped control and readout pulses. Microwave for X/Y rotation and readout is generated by two microwave sources (R\&S SMB100A) and then mixed into the control pulses by IQ mixers. The readout signal is preamplified by a HEMT (low noise factory, LNF-LNC4\_8C) at 4K stage, followed by a second amplification at room temperature, then down-convertded by an IQ mixer and digitized by a high speed digitizer (AlazarTech, ATS9360). The Qubit states are then decided by the amplitude/phase at the driven frequency of their corresponding readout resonator.}
  \label{xmon_qubit}
\end{figure}
\begin{figure*}[ht!]
		\centering
		\includegraphics[height=0.25\textwidth]{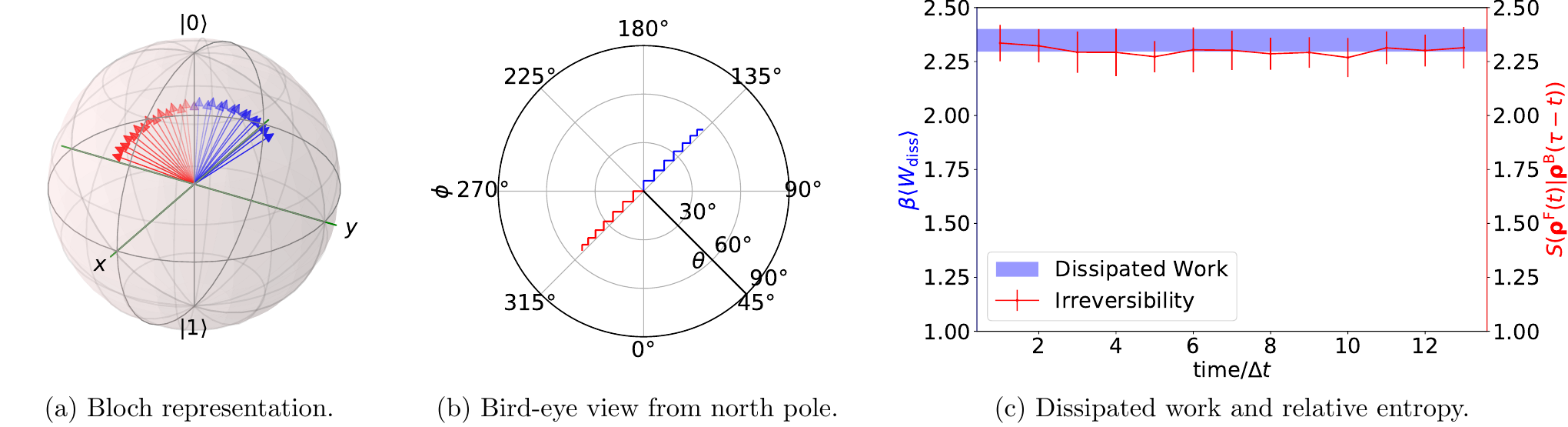}
		\caption{(Color online) Discrete evolution and results. In forward process, the states are consecutively rotated along $x,y$-axis by
angles $\theta=6.22^\circ\pm1.76^\circ$ which last
					$\Delta{t}=16~\mathrm{ns}$. The corresponding backward process is realized as consecutive rotations along negative
					$x,y$-axis in reversed sequence. (a) and (b) depict simulation of externally driven unitary forward
					(red) and backward (blue) processes. In representation (a) and (b), time evolution starts from the north pole,
the $x,y$-axis rotations are consecutively interchanged, shown as zig-zag line in (b). Experimentally,
time evolution starting from south pole is also implemented.
(c) is experimental result of dissipated work and irreversibility
					depicted by relative entropy, showing left-hand-side and right-hand-side
of relation (\ref{time_assymetry_relativeEntr}), respectively. Note that height of the blue strip represents double  standard deviation
					of dissipated work while its center is the average dissipation.
The experiment demonstrates that relation (\ref{time_assymetry_relativeEntr}) is satisfied.}
		\label{xy_rotation_simulation+RelEntExperiment}
\end{figure*}
	\begin{figure}[hb!]
		\includegraphics[height=0.23\textwidth]{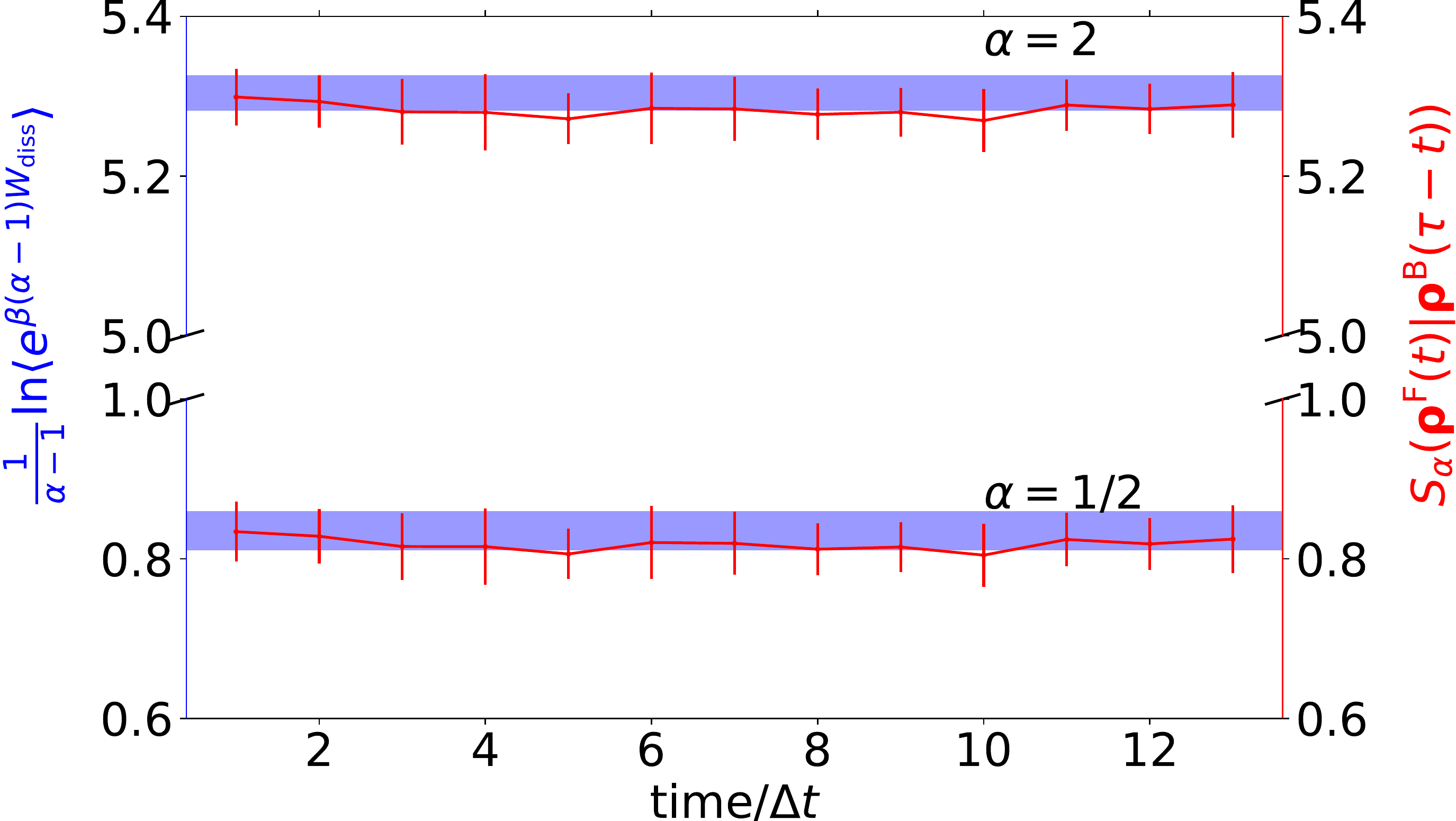}
		\caption{(Color online) Relative R\'enyi entropy of orders $\alpha=1/2,2$.
They are shown down and upper in the figure. Similar representations are adopted as
			that in Fig.~\ref{xy_rotation_simulation+RelEntExperiment}(c).
Both sides of equation (\ref{time_assymetry_relatiRenyiEntropy}) are presented, they agree well with each other.}
		\label{renyiEnt_Dissipation_XY}
	\end{figure}
	\begin{figure*}[ht!]
				\centering
				\includegraphics[height=0.25\textwidth]{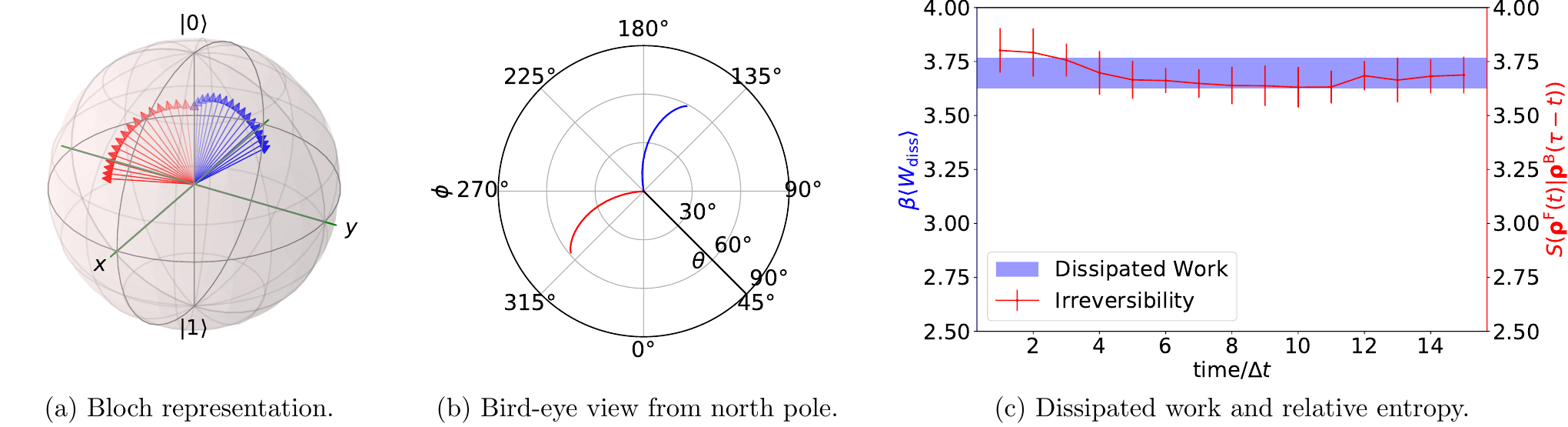}
				\caption{(Color online) Continuous evolution and results. Same conventions are adopted as that in Fig.~\ref{xy_rotation_simulation+RelEntExperiment}.
						 The difference is that the driven protocol is much smooth.
Both (a) and (b) show the forward and backward processes starting from the north pole. In experiment, evolution starting from
south pole is also implemented.
(c) shows both sides of relation (\ref{time_assymetry_relativeEntr}), they
agree with each other.}
					\label{rotatedAxis_rotation_simulation+RelEntExperiment}
	\end{figure*}

	\begin{figure}[hb!]
		\includegraphics[height=0.25\textwidth]{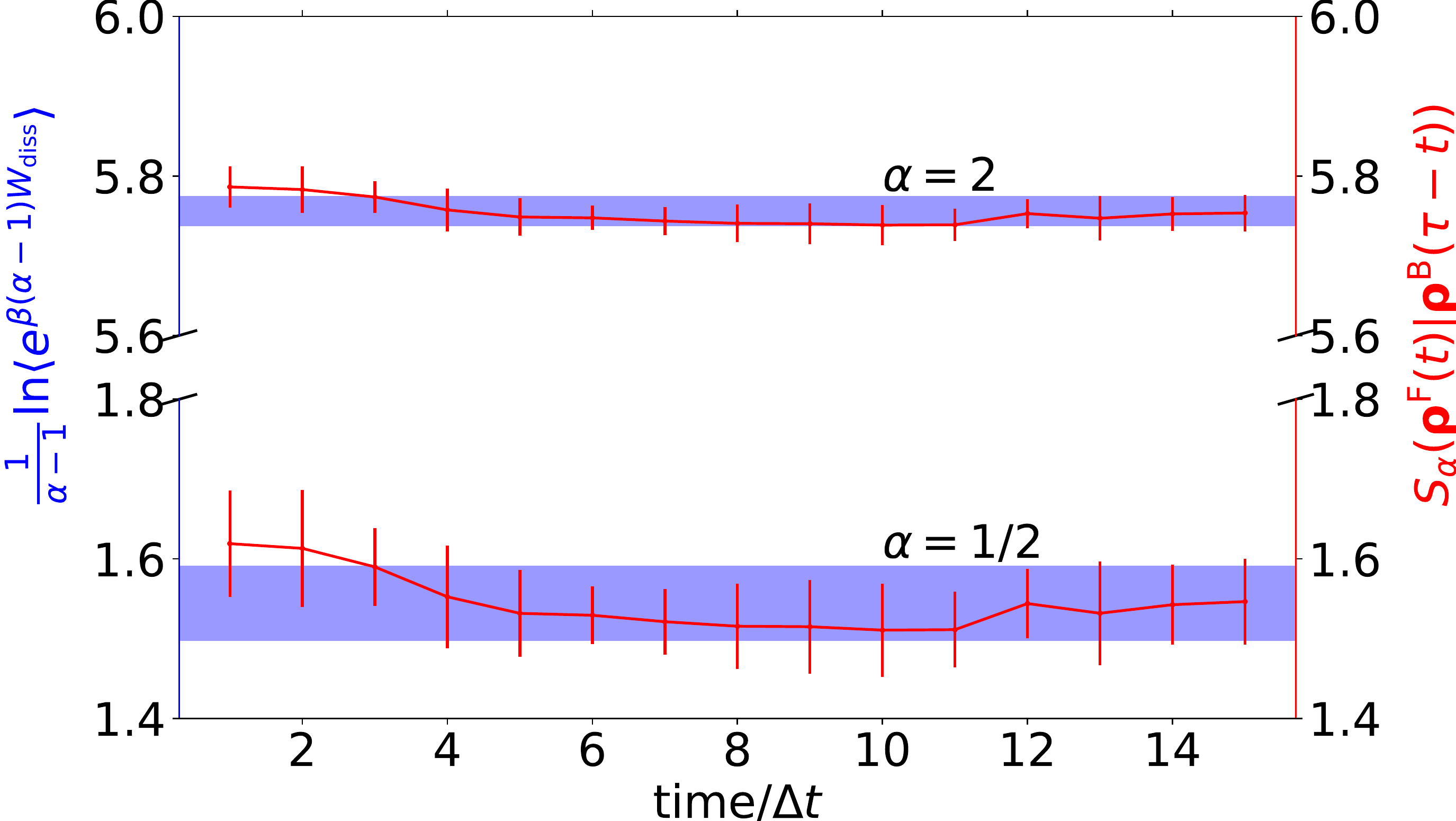}
		\caption{(Color online) Relative R\'enyi entropy of orders $\alpha=1/2, 2$. They are presented
down and upper in the figure. Similar conventions are adopted as that in Fig.~\ref{xy_rotation_simulation+RelEntExperiment}(c).
It is shown that relation (\ref{time_assymetry_relatiRenyiEntropy}) is satisfied.}
		\label{renyiEnt_Dissipation_rotatingAxis}
	\end{figure}

\emph{Experimental setup.}--- Here we report an experimental verification of irreversibility and dissipation
	relations~(\ref{time_assymetry_relativeEntr}) and (\ref{time_assymetry_relatiRenyiEntropy}) in a transmon
	qubit of Xmon variety~\cite{Barends2013,transmon}. A transmon/Xmon can be considered as a nonlinear oscillator with a Josephson junction or a DC-squid acting as its nonlinear inductor. The lowest two energy levels of Xmon are used to implement a qubit. Xmon qubits have improved coherence time~\cite{Barends2013} and can be fully and accurately controlled~\cite{barends2014} with microwave pulses. It can realize qubit state tomography with dispersive quantum non-demolition (QND) measurement~\cite{blais_cavity_2004,Yale2016}. Those features make it a good choice for our experiment. A photography of our device is shown in Fig.\ref{xmon_qubit}(a). It has four Xmon qubits (marked from "qubit 1" to "qubit 4") arranged in a line with nearest-neighbor coupling scheme. The chosen Xmon (qubit 2) was set to its optimal point with a qubit frequency of $\omega_\mathrm{q}/2\pi=6.20240~\mathrm{GHz}{\pm0.02}~\mathrm{MHz}$ and anharmonicity being about $-236~\mathrm{MHz}$. Its readout resonator frequency is at $\omega_\mathrm{r}{\simeq}6.793~\mathrm{GHz}$, which falls in the dispersive regime with an effective dispersive shift $\kappa/2\pi{\simeq}-0.697~\mathrm{MHz}$. Measurements show that its energy relaxation time is $T_1{\simeq}8.327~\mu\mathrm{s}$  and dephasing time is $T_2^*{\simeq}6.813~\mu\mathrm{s}$,
with spin echo Gaussian dephasing time $T_\mathrm{2}^{\rm se}{\simeq}11.722~\mu\mathrm{s}$. The coherence property, although not as good as that of the state-of-art transmon/Xmon fabrications, is believed to be enough for demonstrating the relations, since the longest operation time is about $300~\mathrm{ns}$, which is much shorter than the decoherence time.


\emph{Experimental procedure and result analysis.}--- The qubit is initialized to state $\ket{0}$ or $\ket{1}$ in both the forward and
	backward process. The required Gibbs distribution for thermodynamical equilibrium states would be obtained through assigning proper
	weight as probabilities to these states such that $\beta\omega_0=\beta\omega_\tau=1$ with
	$E_i^{0,\tau}{\equiv}\pm\frac{1}{2}\omega_{0,\tau}$ being energy levels at the beginning and end of the transition,
 and $\hbar=1$. We conduct projective measurements at the end of every unitary process and obtain transition probabilities
	$p^\mathrm{F}(j|i)$ from initial state $\ket{i}$ to final state $\ket{j}$ of the forward evolution. Average dissipated work can be
	computed through these transition probabilities,
	\begin{equation}
		\braket{W_\mathrm{diss}} = \sum_{i,j=0}^1p^0(i)p^\mathrm{F}(j|i)(E_j^\tau-E_i^0) - \Delta{F},
		\label{average_disspated_work}
	\end{equation}
	with $p^0(i){\equiv}e^{-\beta{E_i}}/Z_0$ Gibbs distribution of energy states and
	$Z_{0,\tau}{\equiv}2\cosh(\beta\omega_{0,\tau}/2)$ partition functions.  Other exponential functions of dissipated work can be
	evaluated similarly,
	\begin{equation}
		\braket{e^{\beta(\alpha-1)W_\mathrm{diss}}} = \sum_{i,j=0}^1p^0(i)p^\mathrm{F}(j|i)e^{\beta(\alpha-1)(E_j^\tau-E_i^0 - \Delta{F})}.
		\label{average_exp_W_diss}
	\end{equation}
	To determine irreversibility namely difference between forward process and its reverse quantified by relative entropy
and relative R\'enyi entropy, we implement state tomography on
	carefully chosen time points of the transitions. For every tomography time point $t$ in the forward process,
	state tomography at corresponding instant $\tau-t$ is arranged in the backward procedure. With data collected about these states of
	different instants, we can compute irreversibility described by either relative entropy or relative R\'enyi entropy of order
	$\alpha>0$.

With the help of equations (\ref{average_disspated_work},\ref{average_exp_W_diss}) which will be measured
experimentally, also relative entropy and relative R\'enyi entropy can be obtained by using state tomography, now,
both sides of relations (\ref{time_assymetry_relativeEntr},\ref{time_assymetry_relatiRenyiEntropy}) will be given experimentally.
Then, we can check whether these two equations are satisfied.

The first realization of unitary driven evolution is a sequential quench of a designed sequence of system Hamiltonians,
	\begin{equation*}
		-\omega_0\pmb{\sigma}_z
		\to \underbrace{
		\omega(t)\pmb{\sigma}_x
		\to \omega(t)\pmb{\sigma}_y
		\to \ldots
		}_{14\textrm{ rotations each last for } 16\mathrm{ns}}
		\to-\omega_\tau\pmb{\sigma}_z.
	\end{equation*}
	The $\pmb{\sigma}_{x,y}$-type Hamiltonians are turned on and off consecutively for $14$ times in the form of
	$16~\mathrm{ns}$ modulated Gaussian-type control sequence $\omega(t)$ providing advantage of avoiding leakage error.
State tomography time points are chosen as the ends of each $\Delta{t}$ showtime of $\pmb{\sigma}_{x,y}$
	Hamiltonian. The backward process is set accordingly. We give a depict of simulation of this designed procedure in
	 Fig.~\ref{xy_rotation_simulation+RelEntExperiment}(a) and (b) namely a Bloch sphere representation and a bird-eye view from the
	 north pole $\ket{0}$ of the Bloch sphere, for both forward and backward procedures.

	Experimental raw data obtained in state tomography are of fidelity $F_\mathrm{raw}\ge91.3\%$ compared to that of simulation.
	Given pure states, we can make every density matrix obtained in tomography normalized and thus eliminate errors due to decoherence
	or measurement defects, and obtain fidelity $F_\mathrm{adj}\ge97.7\%$. We compute average and standard deviations of dissipated
	work  according to Eq.(\ref{average_disspated_work}) and of relative entropy using transition probabilities and state-tomography data
	in $25$ repetitions of the experiments. The results are plotted with respect to times as shown in
	Fig.~\ref{xy_rotation_simulation+RelEntExperiment}(c), which shows that $\braket{W_\mathrm{diss}}$ agrees with
	$S(\pmb{\rho}^\mathrm{F}(t)\|\pmb{\rho}^\mathrm{B}(\tau-t))$ perfectly within experimental error as
	predicted~\cite{kawai2007dissipation,parrondo2009entropy}.
	Similarly, we can compute mean value and standard deviation of work exponentials using Eq.(\ref{average_exp_W_diss}) and
	relative R\'enyi entropy of order $\alpha>0$ with the same state tomography data. One can  choose any values of $\alpha>0$,
	for instance $\alpha=1/2$ and $\alpha=2$.
	Our data as shown in Fig.~\ref{renyiEnt_Dissipation_XY} validates the relative R\'enyi entropy and dissipation relation
	(\ref{time_assymetry_relatiRenyiEntropy}) given by Ref.~\cite{wei2017relations} within resolution of our experimental setup.

	Apart from the sequential quenches, we also conduct a ``quasi-smooth'' driven process by varying the qubit Hamiltonian in a much
	smooth manner. System Hamiltonian firstly jumps from $-\omega_0\pmb{\sigma}_z$ to $\pmb{\sigma}_x$ type which change gently in $16$
	consecutive steps to $\pmb{\sigma}_y$ form and finally quench to $-\omega_\tau\pmb{\sigma}_z$
	\begin{equation*}
		-\omega_0\pmb{\sigma}_z \to
		\underbrace{
			\omega(t)\pmb{\sigma}_x
		\to \ldots
		\to \omega(t)\pmb{\sigma}_y
		}_{16\textrm{ steps each last for } 16\mathrm{ns}} \to -\omega_\tau\pmb{\sigma}_z.
	\end{equation*}
	 Modulated Gaussian control sequence $\omega(t)$ is employed in each step. This is a quantum version of using discrete process with
	 small step length $\Delta{t}=16~\mathrm{ns}$ to imitate a continuous process
	 $\pmb{H}(t) = \Omega\lbrack{\pmb{\sigma}_x\cos(\omega_\mathrm{p}t)+\pmb{\sigma}_y\sin(\omega_\mathrm{p}t)}\rbrack$ where
	the rotating frequency $\Omega/2\pi$ is about $\nicefrac{1}{979}~\mathrm{GHz}$ and procession speed of the rotation axis
	$\omega_\mathrm{p}/2\pi=\nicefrac{1}{960}~\mathrm{GHz}$. So system undergoes $16$ rotations along $16$ equally spaced axes on the
	equator of the Bloch sphere by an angle of $\theta=5.88^\circ\pm1.79^\circ$ each time. With the $x$-axis as the starting rotational
	direction, space between neighbour axes is $\phi=6.08^\circ\pm0.80^\circ$. Similarly, we conduct a numerical simulation of this
	smooth discrete quantum transition. Our simulation results are shown in
	Fig.~\ref{rotatedAxis_rotation_simulation+RelEntExperiment}(a) and (b).

	Smoothness of the transition can vividly be seen from the bird-eye view of north pole
	Fig.~\ref{rotatedAxis_rotation_simulation+RelEntExperiment}(b). Experimentally, we conduct the same procedures of initialization, state tomography,
	and transition probability measurements. The tomography time points are chosen to be the end of each $16$ discrete steps.
	Both the temperature and energy levels at $t=0,\tau$ are assumed the same as before. Raw data of tomography shows that the fidelity
	is $F_\mathrm{raw}\ge89.4\%$, which can be improved to $F_\mathrm{adj}\ge94.0\%$ by considering the states
 are pure as shown above. Corresponding experiment data as shown in
	Fig.~\ref{rotatedAxis_rotation_simulation+RelEntExperiment}(c) confirms the relation (\ref{time_assymetry_relativeEntr}) between
	average dissipated work and relative entropy. Also,
	Eq.(\ref{time_assymetry_relatiRenyiEntropy}) can be verified, as shown in Fig.~\ref{renyiEnt_Dissipation_rotatingAxis}.

\emph{Conclusion.}--- We have designed and implemented experiments on the platform of superconducting Xmon qubit
to provide concrete evidences for verification of Eq.(\ref{time_assymetry_relativeEntr}) and
	Eq.(\ref{time_assymetry_relatiRenyiEntropy}). Both of them are quantitative relations connecting dissipated work and irreversibility of
	the thermodynamic process. Our experiment is the first verification of relation (\ref{time_assymetry_relatiRenyiEntropy}) in a
	quantum platform. Those results lay more solid experimental foundation for theory of quantum thermodynamics.

\begin{acknowledgments}
We thank Wuxin Liu and Haohua Wang of Zhejiang University for technical support.
Scipy~\cite{Jones2001,Oliphant2007,Millman2011}, Numpy~\cite{Oliphant2007,VanderWalt2011,Millman2011}, QuTip~\cite{Johansson2012,Johansson2013} and
	Matplotlib~\cite{Thiruvathukal2007} were used in our numerical simulation and illustration preparation.
We thank Franco Nori and Yuxi Liu for useful suggestions.
	This work was supported by Ministry of Science and Technology of China (Grant Nos. 2016YFA0302104 and 2016YFA0300600),
	National Natural Science Foundation of China (Grant Nos. 91536108, 11404386, 11674376, 11374344, 91321208),
Chinese Academy of Sciences (XDPB-0803) and CAS central of excellence in topological quantum computation.
\end{acknowledgments}
\bibliography{Bibliography}
\end{document}